# Thermodynamics approach to near future of civilization

Vladimir Kh. Dobruskin, dobruskn@gmail.com, Beer Yacov, Israel

The purpose of this study is to consider the near future of civilization in the framework of thermodynamics. Kardashev's proposal to evaluate the development of celestial civilizations by the amount of energy they are able to use was adopted to translate the concept of human activity into the language of physics. The discussion is limited to considering the last 500 years of history from the beginning of the scientific and technological revolutions to our immediate future. The application of classical and nonequilibrium thermodynamics is discussed. In the framework of classical thermodynamics, two systems are compared: a) the first is a hypothetical quasi-equilibrium system, Earth without a population. Since biological evolution becomes almost imperceptible for a short period of time, such a system remains in the same pristine state for the entire period; (b) the second is our habitable planet, which is not in equilibrium due to rapid anthropogenic evolution and can be considered as a combination of the first system with human civilization. It is shown that in response to the development of civilization (a) the equilibrium of the hypothetical system is disturbed and processes are initiated aimed at reducing the amount of energy produced, and (b) there is a maximum on the path of civilization development over time. The resistance of nature will continue until a new balance is established, corresponding to a lower level of energy production. The central problem is whether humanity is ready and able to agree on a new balance, otherwise the degradation of our planet can lead to the collapse of civilization.

## 1. Introduction

In recent decades, events in the world around us have changed at breakneck speed. This applies to all aspects of life: climate, the number of natural disasters, outbreaks of new diseases and the coronavirus pandemic, changes in morals and demographics, fierce political struggles, wars and the crisis of the global economy, conflicts of interest between the greens and industry, etc. The media, such as newspapers and television, pour out shocking and depressing news every day. The future of our civilization has been widely discussed in thousands of publications around the world, a complete overview of which is impossible to present here. The problem is discussed by science fiction writers, institutes for the study of future [1, 2, 3], religious figures [4] and many scientists around the world; there are millions of sites on the Internet. There are both gloomy and optimistic forecasts, and the discussions seem likely to go on forever. It can be summed up that the problem affects all of humanity, experts in all fields are concerned about the future of civilization and are trying to find solutions.

In many publications, consideration is based on the implicit assumption that the current course of development of our civilization will be projected into the future and there will be no obstacles to progress. The authors seem to demonstrate their faith in science and its ability to overcome all difficulties. This assumption is challenged by a group of scientists, who studied the possible limit of civilization growth [5, 6, 7]. These studies were initiated by the Club of Rome, which consisted of current and former heads of state and government, diplomats, scientists, economists, and business leaders from around the globe; the Club was created to address the multiple crises facing humanity and the planet. The computer program used by the authors predicted resource depletion by 2040-2050, which would lead to a sudden and uncontrolled decline in both population and industrial capacity. These studies generated debates around the world: some researchers supported the authors, but most sharply criticized them. In regard to this discussion, we adhere to the view of Avery, who stated that the basic thesis—that unlimited economic growth on a finite planet is impossible—was indisputably correct [8]. The esteemed Union of Concerned Scientists issued a warning and called for humanity to take measures to prevent the future collapse of civilization [9, 10].

Our goal here is to consider the near future of civilization from the most general point of view, that is, within the framework of thermodynamics (TD), and to join the call for emergency measures to prevent catastrophic events. We resort to thermodynamics [11, 12] since there are no exceptions to its laws, at least in the part of the universe that determines our existence. Albert Einstein wrote [13] about the classical thermodynamics: "It is the only physical theory of universal content that, I am convinced, will never be overthrown, within the framework of applicability of its basic concepts." Therefore, thanks to such generality, one could hope to get an indisputable forecast for the future. The evolution depends on many phenomena, including those that are difficult or even impossible to predict today (giant earthquakes, huge space disasters). For this reason, we limit ourselves to considering a short period of our history (~500 years) from the beginning of scientific and technological revolutions to the near future, assuming the absence of catastrophes and considering only the consequences of human activity.

The manuscript is structured as follows. After the introduction in section 2, the application of thermodynamics to processes in the biosphere is justified, the translation of the concept of "human activity" into physical language and the preferences for using classical thermodynamics are discussed. Here we will also touch upon the problem of the progress of civilization in the language of information. The reaction of the planet to the human activity, ways and timing of the planet response are discussed in section 3. Finally, the manuscript concludes with a summary (section 4).

## 2. Applicability of thermodynamics to the problem of evolution
### 2.1 Justification of application of thermodynamics

Our task is to predict the reaction of the holistic planet to human activity in the global ecosystem of Earth, that is, in the biosphere. The latter is the shell of a planet, populated by living organisms, which is under their influence and is occupied by the products of their vital activity. The biosphere can be viewed as a huge collection of extremely complex chemical processes. Of course, biochemical processes have their own characteristics: they involve polymer molecules with a complex chemical and spatial structure, reactions occur with the participation of catalysts of the same nature; among biochemical reactions, conjugate processes are widespread, when two or more reactions can be connected (coupled) in such a way that thermodynamically unfavorable reactions and favorable reactions combine to lead the overall process in a favorable direction. For phase transitions, individual and conjugate reactions, the TD methods allow to calculate equilibrium data, temperature dependences, pressure effects, theoretical yields, and so on, that is to predict the effect of changes in the environment on the equilibrium parameters. Chemical thermodynamic [11,12] is one of the main tools of physical chemistry. The results of quantitative TD calculations are often expressed qualitatively in the form of the Le Chatelier principle: if an external influence is exerted on a system that is in a state of *dynamic equilibrium*, then processes will occur in the system aimed at reducing the external influence. As a metaphor, we can say that the Le Chatelier principle reflects the existence of feedback between the system and the environment. It is worth noting that the Le Chatelier principle is generalization of the results of calculations, an integral part of the equilibrium thermodynamics, and not a separate statement; it is applicable only to the dynamic equilibrium states.

For the totality of all the complex processes in the biosphere, it is impossible to make the same quantitative calculations as for individual and conjugate reactions. However, due to the applicability of TD to individual reactions, it can be assumed that it is also applicable to their totality, at least for qualitative predictions. It should be noted that classical thermodynamics was developed before the molecular-kinetic theory and does not require knowledge of the details of the molecular mechanisms. In our case, this means that to apply the laws of classical TD to the planet, one does not need to know the mechanism of processes in the biosphere.

**2.2 The Kardashev Scale**

In 1964, the astrophysicist Nikolai Kardashev, considering hypothetical celestial civilizations, proposed to evaluate their development by the amount of energy, $E$, that they can use [14]. This idea of Kardashev seems acceptable for translating the concept of "human activity" into the language of physics: no human activity is possible without the use of energy. Moreover, humanity produces energy to meet its needs for real goods and services. Perhaps to some, the Kardashian approach may seem at first glance unconvincing, devoid of humanitarian content and reducing all the diversity of human activity to a "soulless" amount of energy used; such a "disregard for details" is universal for the language of physics: it is enough to remember that a movement of an imaginary material point can reflect the movement of a car, a projectile, a planet, a rocket, etc. and geometric points, lines, and shapes, model real material objects. The Kardashev scale opens up the possibility of applying thermodynamics to study the development of civilization.

There is another competing concept of describing the progress of civilization in the language of physics. In this regard, it is necessary to focus on the role of information. Harari notes [15] that we live in a world created by our imagination, we are surrounded by concepts that do not exist in nature: culture, ideology, morality, political parties, states, money, nations, stocks, law, religions, science, human rights, etc. – all these are manifestations of our cognitive abilities, and human activity is completely determined by the mental activity. To characterize the latter, the concept of "information" is often used and changes in the world are explained by the explosive growth of information.

Due to the breadth of this concept, there seems to be no universal definition of the term "information" [16]. One might recall the famous statement of the "father" of cybernetics N. Wiener [17]: "Information is information, not matter and not energy". Wiener believed that information is akin to such categories as movement, life, consciousness. Information stored in libraries, museums, computers, etc. has no effect on nature as long as it is not used by people. Since human activity is impossible without information, the expressions "explosive growth of information" and "explosive growth of energy produced" are in unambiguous agreement, and it is incorrect to apply them as independent causes of events. In fact, the use of these concepts looks like a description of the same phenomenon in different languages. The question arises: "Which language is preferable for the quantitative description of evolution?" We believe that a quantitative description in the language of information is currently impossible.

Such attempts are made by introducing information entropy (Shannon entropy) to thermodynamic relations. We believe that it is a mistake and below present the arguments in favor of our opinion: There are three definitions of entropy: (a) the classical one of Clausius [11], $dS_c$, (b) the definition of Boltzmann [11], $S_B$, and (c) the entropy of Shannon, $S_{Sh}$, which is associated with the amount of information, $I$, [18-20]. The notations of $S_B$ and $S_{Sh}$ came from statistical mechanics. The concept of information entropy refers to the theory of information transmission, and has no connection with the origin and the value of information. On the other hand, the mental activity of people, as well as the new information (knowledge), is the result of the work of consciousness that obviously has nothing to do with statistic mechanics (the molecular-kinetic theory), and, therefore, the Shannon entropy also has nothing to do with the appearance of new information and its impact on human society: it is impossible to quantify such manifestations of people's cognitive abilities as culture, morality, religion, etc. Consider the following example. Let's (a) write on one sheet of paper the Ten Commandments, and on the other sheet of paper -a meaningless text of the same length from the tabloid, and (b) count the amount of information in both texts. Examples of such calculation are given in the literature [21]. We will find that the amounts of information are about the same, while the impact of the Ten Commandments on society is huge, but the impact of the tabloid text is most likely close to zero. Hence, it makes no sense to use information entropy in the thermodynamic consideration of evolution. Part of the problem is that three abovementioned definitions have the same name- entropy. Since there is no quantitative theory of the impact of information on humanity (not to be confused with the theory of communication!), then, at least at the moment, it is impossible to quantify anthropogenic activity in terms of information, and there is only one option left - to evaluate it on the basis of the Kardashev scale.

Let's look at what the energy of civilization is. Consider two thermodynamics systems: the first hypothetical system is Earth without human population (a land untouched by civilization), and the second is our actual inhabited planet. Then, the energy, $\Delta E$, brought by civilization is the excess energy, that is, the difference between the energies consumed by the real and hypothetical systems:

$$\Delta E = E_{ap} - E_{hp} \qquad (Eq.1)$$

were $E_{ap}$ and $E_{hp}$ are the energies consumption of the actual and hypothetical planets, respectively.

According to the origin, there are two types of energy sources-renewable and non-renewable. Renewable energy originates from the current energy of solar radiation; it does not depend on whether the planet is inhabited or not. In the case of planet without people, most of the current solar energy is spent on maintaining the state of the planet, a small part is stored in natural storage-deposits. In the case of our inhabited planet, a small fraction of energy is used in wind-, solar-, tidal- and hydroelectric power plants to produce useful work for people, the latter in the course of its use will also turn into heat. For example, an electric car carries a load, but all useful energy will be spent on overcoming friction and eventually also turn into heat. Therefore, renewable energy, which serves also as the energy source of life, has approximately the same effect both on a planet without people and on a real planet. These values disappear when calculating the difference $\Delta E$ (Eq.1).

In the case of non-renewable energy, the situation is different: it is also solar energy, but accumulated over hundreds of millions of years by an uninhabited planet; the stored energy is released by humans through the rapid burning of fossil fuels. All other components of the Earth energy balance also do not depend on whether the planet is inhabited or not and disappear when calculating the difference $\Delta E$. Only one component remains- the energy produced by civilization, that is non-renewable, mainly, energy of fossil burning. This fact radically simplifies the calculations, since the latter is reported in official data.

The growth of energy consumption $E$ began after the appearance of *Homo sapiens* about 50,000 years ago. We can assume that the destructive effect of the fuel combustion energy is summed up over time, *i.e.*, the integral

$$\Delta E = \int_{-50000}^{t} E(t)dt \qquad (2)$$

rather than the current value of *E*, has a destabilizing effect on the environment, where *t* is time. The value of the integral and, hence, the destabilizing effect, increase with time. During most of this period, the state of nature was little variable, and only after the beginning of the scientific and technological revolution, the integral (equation 2) begins to increase rapidly. In our earlier work we provided the example of Δ*E* calculation [22].

It should be emphasized that on a planetary scale, all the energy produced is consumed and nothing is saved for the future. Consequently, the amounts of energy produced and used are equal *on a global scale*, but not in the case of individual countries, which can be either exporters or importers of energy. Because of the quantitative equality, the terms "energy production" and "energy consumption" on a global scale sometimes will be used interchangeably. It is done for convenience, depending on the circumstances, and not as a result of confusion.

Kleidon suggests evaluating the activity of people by the amount of free energy used by humanity [23]. However, to assess the damage to nature, in our opinion, it is necessary to use the total, and not free energy. If fossil fuels are burned with zero efficiency, the damage to our planet as a whole would be the same as if they were used efficiently. The free energy defines the limit efficiency of energy use *for the people*, but the damage *to our holistic planet* is defined by the total energy.

## 2.3 Biological and historical evolutions

Biological evolution is usually considered within the framework of Darwin's theory. The latter assumes that when the environment changes, only organisms with genetic mutations that allow them to exist in the new environment survive. Biological evolution, with the exception of protozoan evolution, is a slow process that takes thousands of years or more before significant changes occur, and within a short period of time, biological evolution becomes almost imperceptible. Harari points out [15] that with the emergence of Homo sapiens, who in the course of the cognitive revolution learned to quickly correct their behavior and pass on new skills to the following generations, even without genetic mutations or changes in the environment, the historical evolution begins. Unlike slow biological evolution, a rapid historical evolution can change the living conditions within a single generation of people. By the term "historical evolution", we mean the change in time of the totality of processes in human society and the environment, which are generated due to the human mental activity. It should be noted that other authors [24] give a similar description of evolution, calling this process cultural evolution. They also note that the transition from a genetic type to a cultural type of evolution has been made possible thanks to the extraordinary development of the human brain [24].

**2.4 Evolution in the framework of classical and nonequilibrium thermodynamics.**

Any application of the laws of thermodynamics begins with an imaginary division of the universe into a system (area of interest) and the remainder of the universe that lies outside the boundaries of the system - the environment. Since the universe is an isolated system, its entropy, *S*, accordingly to the second law of thermodynamics, can only increase:

$$\Delta S \geq 0 \qquad (3)$$

where equality is possible only if an equilibrium is reached, the state that Clausius called the heat death of the universe. Consider our planet together with the atmosphere as a system, and space as an environment.

It is known that the main impact on our planet is caused by solar radiation. The study of the earth's energy balance between incoming short-wave solar radiation, on the one hand, and outgoing radiation from Earth, on the other, shows [25] that the average annual energy influx into the Earth's system is equal to the return radiation into space. Therefore, on average, the change in the thermal energy of Earth is zero, $q=0$. More detailed information about the energy balance can be found in the original source [25]. In addition, Earth and space practically do not exchange matter: the volume, *V*, and mass of the planet remain almost unchanged. Although there has been an increase in temperature in recent years, the absolute increments are small compared to the average temperature $T=288$ K. For example, over 138 years (1880-2018) of observations [6], the temperature ranged from -0.5 K to +1 K from the average temperature of the twentieth century. Therefore, we can roughly assume that the processes at small time intervals occur at a constant temperature. The planet does not produce work, *A*, in the surrounding space, and in the formulation of the first law of thermodynamics

$$q = \Delta U + A \qquad (4)$$

where $\Delta U$ is the change in the internal energy of the system, all components are equal to zero.

We will confine ourselves to events occurring during a short time interval compared to the period of the planet's existence. It should be emphasized that *this is an important, key limitation*, without which further consideration will be incorrect. The short interval in this case is the last ~500 years of our history from the beginning of scientific and technological revolutions to our near future, that is ~ $10^{-7}$ part from the total Earth history. During this period, the processes occurring on Earth practically did not affect the surrounding space, so we will limit ourselves to discussing changes only on the planet.

In general, the evolution of a thermodynamic system can be considered in the framework of either classical equilibrium TD, or nonequilibrium thermodynamics (NTD). Equilibrium thermodynamics restricts its considerations to processes that have initial and final states of thermodynamic equilibrium; it is not a theory that included the irreversible processes that are responsible for the transformation of one state to another. In contrast, NTD attempts to describe these transformations [26]. In this science, the dependence on time has become quantitative; it is based on the calculation of the rate of increase in entropy, i.e., a dissipation function. Onsager, Prigogine, and others considered open, nonequilibrium but close to equilibrium systems, and postulated linear relationships between the TD parameters; they laid the foundation for a linear NTD [26-30]. In recent years, the third stage of the development of thermodynamics has come - the physics of dissipative systems, i.e., the physics of non-equilibrium processes that are far from equilibrium [28-30]. This branch of thermodynamics is a work in progress. The conclusions drawn from thermodynamics consideration depend on what is recognized by the system: (1) If we focus only on an uninhabited Earth with an atmosphere, then this system is in quasi-equilibrium, which, according to the NTD, is maintained by a constant flow of entropy through it [31]. (2) If we see the same system unchanged for 500-1000 years (see further) and if we are not interested in (a) how it came to be at the initial point and (b) the mechanism of its transition to the final point, we can consider the system during this interval as an equilibrium system governed by the laws of classical thermodynamics. (3) A habitable planet for any period is a non-equilibrium system due to the development of civilization, and finally, (4) a living organism (up to its death) is a non-equilibrium open system that exchanges matter and energy with the environment.

What we really want to do is answer the question: "How will anthropogenic activity change the balance in a hypothetical, ideal nature?" The direct way to find the answer lies within the framework of equilibrium thermodynamics. Another reason for its application is the following. The events of the last ~ 500 years (~$10^{-7}$ of the total planet history) cover an explosive, nonlinear, period of civilization development; this is a period where, strictly speaking, the linear approximation of NTD does not work. To apply NTD to very complex processes on Earth, simplifying assumptions must be made that will devaluate the result. Unlike NTD, classical thermodynamics describes processes, for which only the start and end points are important, even if the path between them involves an explosion.[32]

## 3. The reaction of the planet to the progress of civilization. Discussion

**3.1 Shift equilibrium as a response of the thermodynamics system**.

We will continue to consider the hypothetical system, the Earth in the absence of people, in the specified short historical period. The system is a dynamic equilibrium, when at the macro level, on average for a year and over the entire surface, no changes are visible, but at the "molecular" micro level, the processes continue: the planet rotates around its axis and the sun, winds blow, water evaporates, seasons change, it rains, organisms are born and die, etc. But for an external observer who does not distinguish "small" details, the state of the planet on average seems to be non-changeable, i.e., at the macro level, the unhabituated planet is in equilibrium. The system is similar to the simplest one: the macro-volume of the gas is in equilibrium, while the gas molecules are in chaotic motion. The primeval nature of the hypothetical world is the desired goal of humanity. It is worth emphasizing once again that the entire planet, and not its individual components, is in balance. As has been said, a living organism is never in balance; for it, balance is death. The question of what life is in terms of TD was studied by Schrodinger [33], and investigated by methods of nonequilibrium TD [28-30]. In this paper, open systems, in particular living organisms, as a separate subsystem of Earth, are not considered.

In the absence of a population, the hypothetical planet would have remained in the same state until now. Such equilibrium system is a typical object studied by classical thermodynamics. It obeys the Le Chatelier principle, which states that if a system in a state of dynamic equilibrium is affected by an energy disturbance, then processes are initiated aimed at countering this disturbance. The more $\Delta E$, the stronger the reaction. The resistance of the planet will be aimed at reducing the amount of energy produced. Our actual planet is seen as a combination of a hypothetical system, the planet without a population, and a human civilization. It has not been in equilibrium: anthropogenic evolution continued and accelerated. It can be considered as an indicator of where the displacement of the equilibrium of the hypothetical planet leads to. One may see that, on the one hand, humanity develops civilization, and on the other - simultaneously destroys it. Hence, there is a maximum in the way of the development of civilization over time. One can prove the existence of a maximum on the curve $\Delta E(t)$, that is, a turning point, by simple reasoning. It is obvious that in the absence of resistance to the destruction of nature, the existence of people will become impossible and $\Delta E(t)$ will fall to zero. Hence, the curve that starts from zero at the emergence of Homo sapiens, grows and returns to zero, passes through the maximum.

If a global agreement is reached to reduce energy production, the entire system will come to a new equilibrium; otherwise, it will lead to the disappearance of civilization. Obviously, the turning point between growth and decline of civilization is not just a point on a geometric curve, but a period of time when the first signs of an approaching catastrophe begin to appear. Next, we need to answer two questions: "How can the planet's resistance be realized? When will the tipping point come?" The classical thermodynamics can not give answers to these questions.

**3.2 How could the planet's response be realized?**

The question can be reformulated: "How could the laws of nature force a decrease of the sapiens activity?" The laws of nature do not directly affect the energy production: nature has nothing like army or police. The only possible counteraction is the evolution of the environment towards the deterioration of human living conditions, which would lead to a reduction in the population, that is, to a reduction in the number of energy producers. The latter can be achieved in two ways: by direct change in people's living conditions, or by an indirect, "humanitarian" response. A climate change and epidemics are the direct, formidable weapons in nature`s arsenal. A probability of the epidemics increases with increasing the number of virus mutations in a contaminated environment. Global climate change is generally recognized as the main threat to our civilization; many other risks, such as increasing natural disasters, diseases and epidemic, deforestation, the emergence of new deserts, lack of drinking water, etc., are associated with the climate change [9, 10].

The indirect response can be traced in the chain of historical events in human society. Harari stresses [15] that historical evolution (moving forward) occurs due to the implementation of all the many manifestations of cognitive abilities in culture, ideology, morality, religion, social and economic relations, science, law, etc. As a result of thousands of years of development of these abilities and interaction between them, the modern society was formed. Thus, combining the approaches of Kardashev and Harari, we come to the conclusion that the level of development of civilization is determined by economic growth, as well as humanitarian and political processes.

It seems highly probable that the development of civilization, manifested in the form of an increase in energy consumption, goes through a maximum and, in the absence of restrictive agreements, the destruction of the environment and civilization will follow. This raises the question of what is happening in the humanitarian sphere during this period. It is clear that cognitive abilities, especially the development of science, were the vanguard of progress, and only after they reached a certain level, economic growth began. These are interrelated phenomena, but cognitive progress precedes economic growth. Consequently, as the tipping point approaches, changes may first be expected to be felt in the avant-garde humanitarian sphere: in morality and culture, the emergence of new art, changes in sexual behavior, the rejection of old values, attempts to replace the old conservative ideology with a new one, etc. It can be expected that these phenomena will be associated with the general trend of population decline. Here are two examples of such phenomena: a) increasing energy consumption and the well-being of the population facilitates access to education, medicine, changes priorities, reduces the birth rate, leads to demographic changes and, as a result, to a reduction in the number of local people; (b) changing sexual behavior also leads to a decrease in the birth rate and population. Inevitably we will see escalating conflicts associated with the unequal distribution of civilization's achievements and struggle for resources. In practice, these are manifested in a sharp political struggle and emergence of new threats.

**3. 3. When can the response be realized?**

It is known that classical thermodynamics does not consider time as an independent variable and, therefore, can only indicate the direction of change, but not the time when it takes place. Thus, you need to find other ways to estimate the time. In our case, it seems reasonable to use the conclusions followed from the study of the limits of the growth of civilization [5-7]. The authors predicted that the planet will reach the limit of growth by about 2040-2050 due to the rapid depletion of natural resources and increasing environmental pollution, and after this time, the decline in population and energy production will become inevitable. Taking into account the conclusions drawn from the thermodynamic consideration, it seems that uncontrolled destruction of the habitat can occur even before the most important non-renewable energy sources are exhausted. It is reasonable to believe that the planet is close to a turning point, hence the evidences in the growing tension in all areas that determine the existence of people, related to climate change as well as to humanitarian changes in morality, the decline in the influence of religion, irreconcilable conflicts between conservatives and liberals, insoluble contradictions between states, the threat of nuclear war, and many others.

## 3. 4. Hypothesis of Lovelock

The author would like to note that there is a rather unorthodox hypothesis of Lovelock [34], the conclusion of which probably partly resembles the conclusion from this article that the planet resists change. Lovelock assumed that from the moment of the origin of life, the planet should be considered as a single system consisting of the atmosphere, lithosphere, hydrosphere and biosphere, where each of its components affects the development of other components, and the evolution of life is closely related to the evolution of its physical environment throughout Earth. But Lovelock came to the radical conclusion that Earth and the planet's global ecosystem behave like a biological superorganism, rather than an inanimate object; he argued that those species that negatively affect the environment and make it less suitable for offspring will eventually be driven off the planet, and called his hypothesis the "Gaia theory" in honor of the ancient Greek goddess of the Earth.

From the very moment of the appearance of Lovelock's hypothesis, it was criticized. Some of the fiercest opponents of the Gaia theory were Darwinian biologists [35]. They argued that the cooperation required for the self-regulation of Earth could never be combined with the competition required for natural selection. Besides, the name taken from mythology also caused discontent, and the identification of Earth with a superorganism contradicted the basic concepts of science. Although, on one hand, the name "Gaia theory" seems more like a publicity stunt to draw attention to environmental issues, the Lovelock hypothesis, on the other hand, does indeed resemble an ancient archaic religion. It is not necessary to fantasize that Earth is a biological superorganism to justify the existence of feedback: the latter follows from the Le Chatelet principle. Although the theory was rejected, this does not prevent it from continuing to excite the imagination and attract many supporters around the world.

## 3.5 Is the planet' response applicable to its separate part?

Above, we concluded that conditions may arise that will lead to a reduction in the population. Let's mentally divide the entire population into several parts according to the standard of living, such as (a) the "golden billion" states, (b) China, (c) India, (d) Russia and neighboring states, (e) South and Central America and (f) Africa. We want to know if the conclusion is true for separate parts of the total population. The exact answer is no, since all the subsystems are interconnected. Although the major crisis will affect the whole mankind, it still seems that the first victim may be the golden billion. You can draw such a simple analogy. Considering themselves the vanguard of

14humanity, these countries rush up to the top of civilization, not suspecting that there is a chasm on the other side of the summit. Throughout history, these states have contributed the most to the pollution of the planet, separated themselves from the rest of the world and for a long time have exploited the resources of the other countries. Looking back, in all periods of history, there were groups of people (Empires) who considered themselves the chosen people, the supreme race, the exclusive nation, tried to teach and subdue the world around them, and suffered a historic defeat. This course of historical events looks like a manifestation of Darwinism in history. The difference from the animal world is that people, due to their cognitive abilities and dominant position in nature, destroy their own habitat, even without other contributing factors. It may well be that this effect of civilization is determined by the laws of physics, and not of our free will.

## 4. Conclusions

The main content of the proposed article can be summarized as follows. Within the framework of thermodynamics, it is considered how human activity in 500 years turns a hypothetical calm planet in balance with the environment into a place where processes aimed at the destruction of humanity are initiated. The discussion is conducted within the framework of classical, but not nonequilibrium, thermodynamics.

The conclusion that the planet will resist human activity, deduced from thermodynamic considerations, seems indisputable. Of course, it is valid only if there were no mistakes when applying TD. The resistance will continue until a new balance is established that corresponds to an acceptable lower level of energy production. The latter will happen if humanity is willing and able to agree on a new balance. This is a central problem that depends on political decisions, otherwise degradation can lead to disaster. Given what is at stake, it is to be hoped that the sooner the better, meaningful measures will be taken to achieve a reasonable balance.

**Acknowledgement**

The author would like to thank Arkady Rutkovskiy for useful conversations and valuable comments.

**References.**